\documentclass[preprint2,twoside]{hwo}

\usepackage{float}

\newcommand{\uas}{\ensuremath{~\mu\text{as }}}
\newcommand{\um}{\ensuremath{~\mu\text{m }}}

\input{hwo.h}

\begin{document}

\title{\textbf{\LARGE Experimental tests of the calibration of high precision differential astrometry for HWO}}
\author {\textbf{\large Manon Lizzana$^{1,2,3}$, Fabien Malbet$^1$, Alain Leger$^4$, Fabrice Pancher$^1$, Sébastien Soler$^1$, Hugo Rousset$^1$, Thierry Lepine$^5$,  Julien Michelot$^3$, Yahya Er-Rahmaouy$^1$, Youssef Bakka$^1$}}

\affil{$^1$\small\it Institut de Planétologie et d'Astrophysique de Grenoble, observatoire des sciences de l'univers de Grenoble, Université Grenoble Alpes, Centre National de la Recherche Scientifique, France}
\affil{$^2$\small\it Centre National d'Études Spatiale, France}
\affil{$^3$\small\it Pyxalis, Moirans, France}
\affil{$^4$\small\it Institut d'astrophysique spatiale, Institut National des Sciences de l'Univers, Université Paris-Saclay, Centre National de la Recherche Scientifique, Centre National d’Études Spatiales, France}
\affil{$^5$\small\it Laboratoire Hubert Curien, Institut d'Optique Graduate School, Université Jean Monnet - Saint-Etienne, Centre National de la Recherche Scientifique, France}

\begin{abstract}
Many different scientific applications require sub-micro arcsecond precision astrometry, including researching rocky exoplanets in the vicinity of the Sun and studying dark matter. The Habitable Worlds Observatory (HWO) is a promising candidate to carry an astrometric instrument because it provides a stable, space-based telescope with a large aperture, which allows faint sources and small displacements to be observed. This paper presents the characterization of an appropriate detector for an astrometric instrument: the 46Mpx Gigapyx from Pyxalis. Moreover it explains the implementation of a testbed enabling interferometric characterization of pixel positions. Finally, the paper introduces a method for calibrating the telescope's optical distortion. This method was implemented in simulation and tested thanks to an optical bench developed at IPAG in France.\\
\end{abstract}

\vspace{2cm}

\section{Introduction}

Differential astrometry is the measurement of the position and motion of a target object relative to the stars in its field of view (FOV). This technique increases the precision on a pointed object with sit-and stare observations.\\

Astrometry is a subject of interest for many science cases \citep{2021ExA....51..845M}. First, astrometric measurements are needed to study dark matter (DM): the nature of DM,  the outer Milky Way halo, the DM subhalos in the galactic disc, the substructure population of the DM halo. In addition astrometry may allow to explore planetary systems in the vicinity of the Sun and particularly rocky exoplanets in the habitable zone of nearby solar-type systems. Finally, astrometry can help to study compact objects.\\

The strongest constraint is imposed by the exoplanet research, the end of the mission required precision is $0.3$~\uas corresponding to the signal produced by an exo-Earth orbiting a Sun-like star $10$~pc away. HWO's focal length is currently estimated at $130$~m, so $0.3$~\uas corresponds to approximately $4.10^{-5}$~px if we assume a detector with a pixel size of $4.4$\um (see section~\ref{characterization}), or  $2.10^{-5}$~px with a pixel size of $10$\um. Thus high-precision calibrations of the detector and the optical distortion are needed \citep{Malbet_2024}.\\

\section{Proposed focal plane concept}
\label{characterization}

\begin{figure*}[!h]
    \centering
    \includegraphics[width=0.45\textwidth]{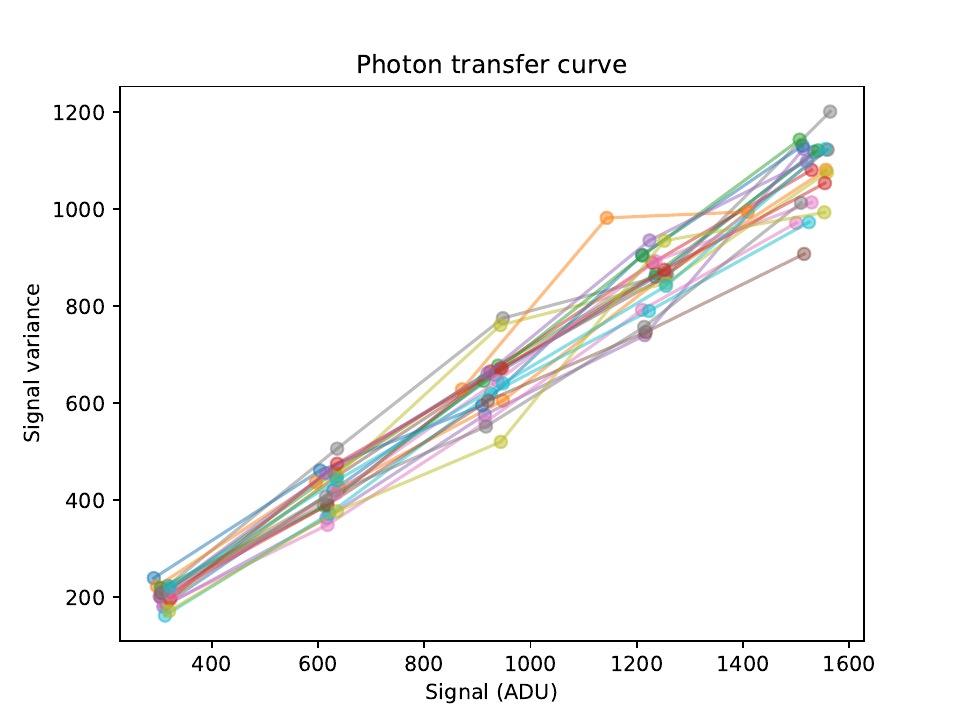}
    \includegraphics[width=0.45\textwidth]{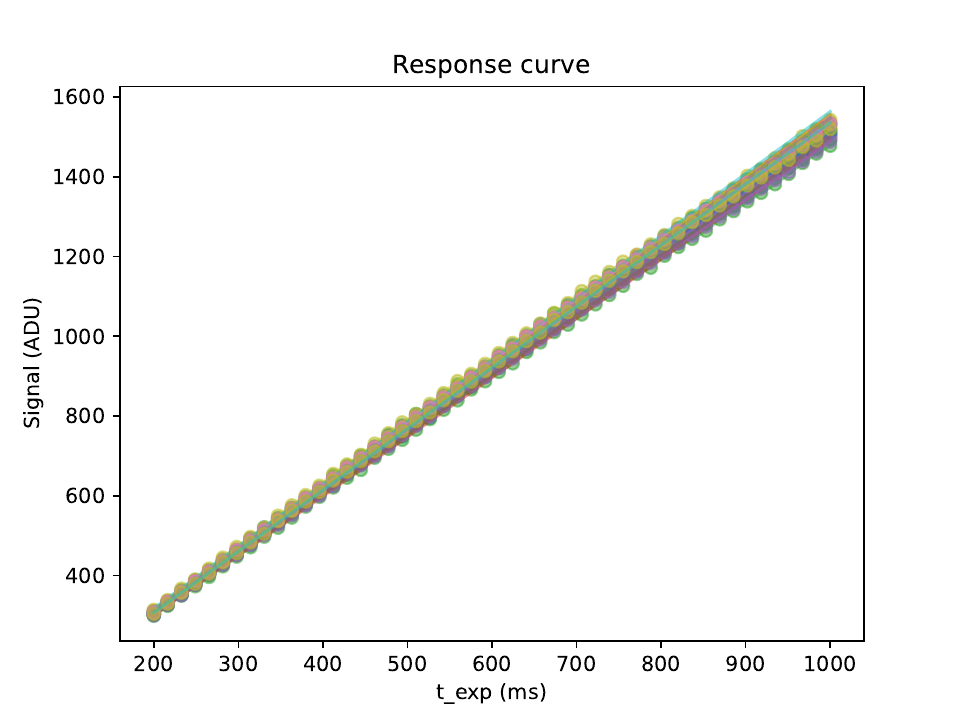}
    \includegraphics[width=0.45\textwidth]{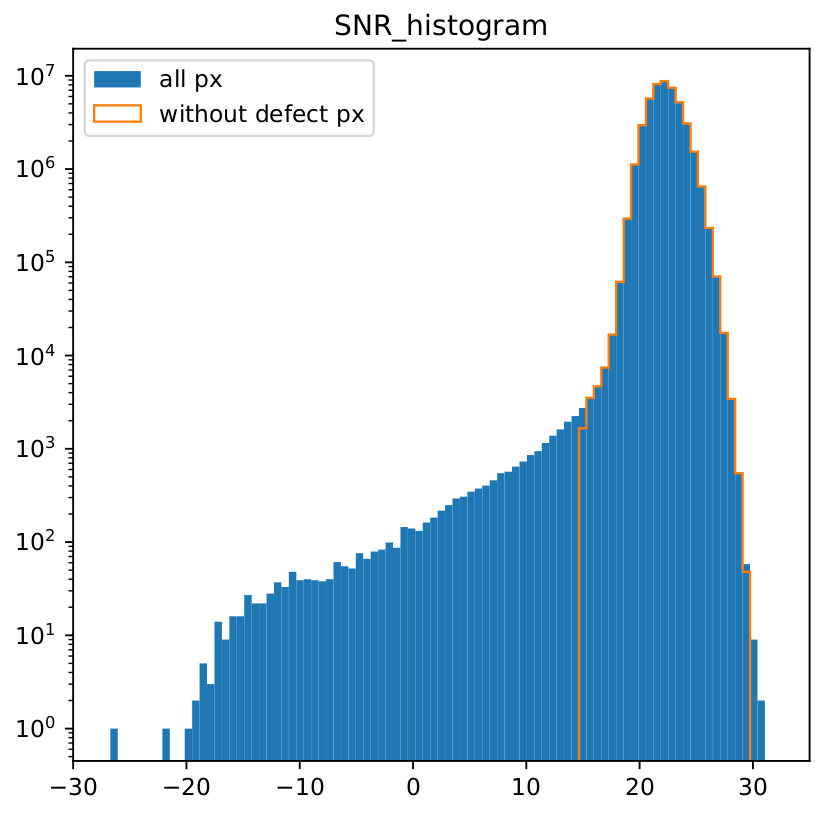}
    \includegraphics[width=0.45\textwidth]{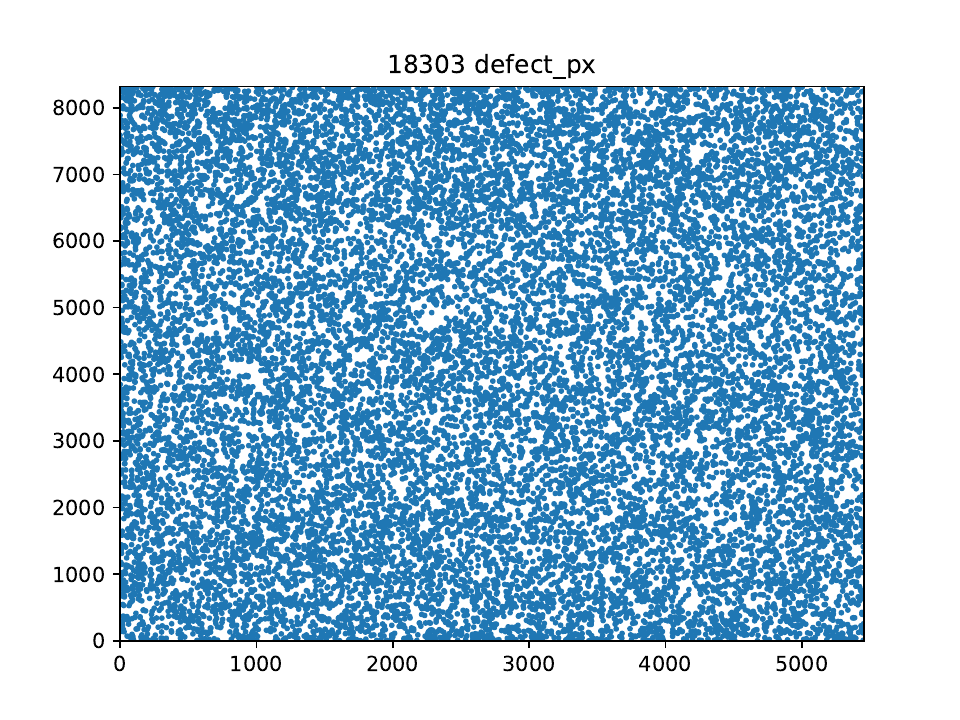}
    \caption{Characterization of the Gigapyx 46Mpx detector from Pyxalis. Top left : photon transfer curve; top right : response curve; bottom left : signal to noise histogram; top right : bad pixel map}
    \label{fig:carac_results}
\end{figure*}

The detector used in this paper is the Gigapyx 46Mpx from Pyxalis (Moirans, France). It is a CMOS sensor with $8320\times5436$ pixels. The pixel size is $4.4$\um, leading to a total size of $24\times26$~mm. It is sensitive to visible wavelengths. This kind of detectors allows to be Nyquist sampled for $\lambda=400$~nm: $f\frac{\lambda}{D} \simeq 8.7~\mu\mathrm{m} \simeq 2$~px, with a focal length $f=130$~m and a diameter $D=6$~m.\\

A global characterization is needed to understand how the detector works and check it meets the requirements. We evaluated the linearity, the dark current, the readout noise, the gain, the pixel response non-uniformity, and the bad pixels. The photon transfer curve, the response curve, the signal to noise histogram and the bad pixel map are pictured in Fig.~\ref{fig:carac_results} and the main results are summed up in Table~\ref{tabular}. Theses figures adhere to the standard for characterization of images, sensors and cameras \cite{EMVA}. The measured performances are compliant with the Pyxalis figures and with the specifications for an astrometric instrument onboard.\\

\begin{table}[!h]
\caption{46M Gigapyx characterization, comparison between IPAG and Pyxalis}
\smallskip
{\small
\begin{tabular}{lll}
& Pyxalis & IPAG\\
\tableline
read out noise & 1.6 e$^-$ & 0.8 e$^-$\\
linearity & $<$3\% & 0.5\%\\
bad pixels & 0.1\% & 0.04 \%\\
response non uniformity & 2\% & 0.8\%\\
full well (low gain) & 50 000 e$^-$\\
\label{tabular}
\end{tabular}}
\end{table}

\begin{figure}[!h]
    \centering
    \includegraphics[width=\columnwidth]{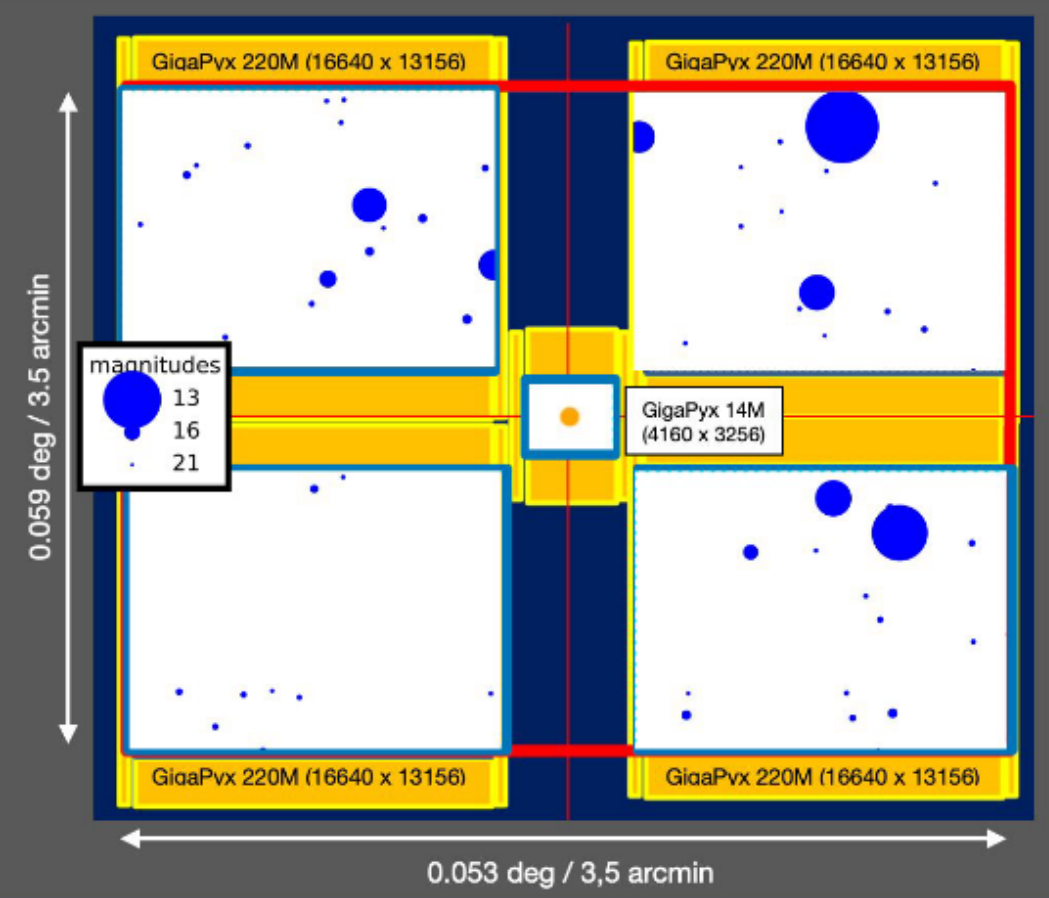}
    \caption{Example of focal plane assembly with four 220M detectors and one 14M}
    \label{fig:focal_plane_assembly}
\end{figure}

The Gigapyx 46Mpx is an interesting prototype as a proof of concept, but it does not cover the whole FOV of the HWO's wide-field imager (WFI). The WFI's FOV is planned to be approximately $3~\mathrm{arcmin}\times 4~\mathrm{arcmin}$ and the focal length about $130$~m which leads to approximately $25000\times35000$~px or about $1$~Gpx. Moreover Pyxalis proposed a 220Mpx detector similar to the 46Mpx we use. Then we proposed a focal plane assembly showed in Fig.~\ref{fig:focal_plane_assembly} \citep[see][for more details]{proceeding_jerome}.\\

\section{Interferometric calibration of the pixel centroid position}

In a theoretical detector the rows and columns are well aligned, however in a real detector the centroids are misaligned due to the fine pixel structure, quantum efficiency local variations ... These misalignments can be as much as $5.10^{-2}$~px or $5.10^{-3}$~px, so calibration is necessary to achieve a final accuracy of $5.10^{-6}$~px.\\

The calibration we propose is based on an interferometric pattern. Two fibers form Young's fringes on the detector and a phase modulator make the fringes scroll along the detector over time. Each pixel observes the modulation, which provides the position of the centroids.\\

\begin{figure*}[!h]
    \centering
    \includegraphics[width=0.9\textwidth]{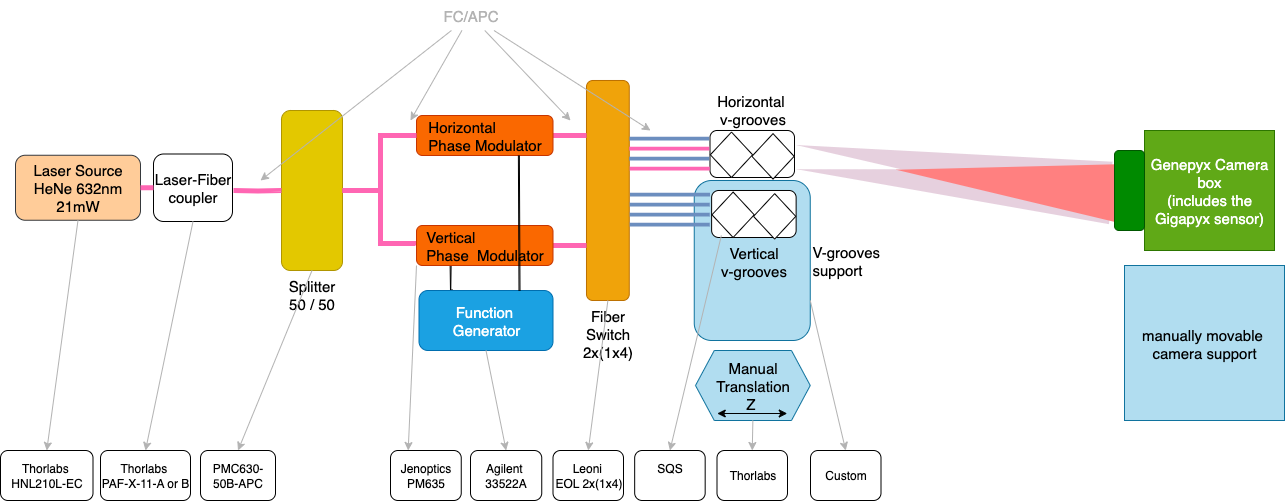}
    \caption{Scheme of the interferometric calibration testbed at IPAG, France \citep[see][]{pancher2024laboratorycharacterisationbenchhigh}}
    \label{fig:banc_calib}
\end{figure*}

Two proofs on concept have been already completed. The first by \cite{2016A&A...595A.108C} achieved $6.10^{-5}$~px and the second one by \cite{2023PASP..135g4502S} achieved $3.10^{-5}$~px. Now, performance must be validated on large matrices, for this purpose a testbed has been set up at IPAG (France), see Fig.~\ref{fig:banc_calib}. An HeNe laser source ($632$~nm) is divided into two fibers by a splitter, a  LiNbO3 electro-optics modulator from Jenoptik modulates the phase between the two beams. Then a fiber switch allows to chose the direction and the baseline between the outputs. $40$~cm away the genepyx records the diffraction pattern.\\

\begin{figure*}[!h]
    \centering
    \includegraphics[width=0.45\textwidth]{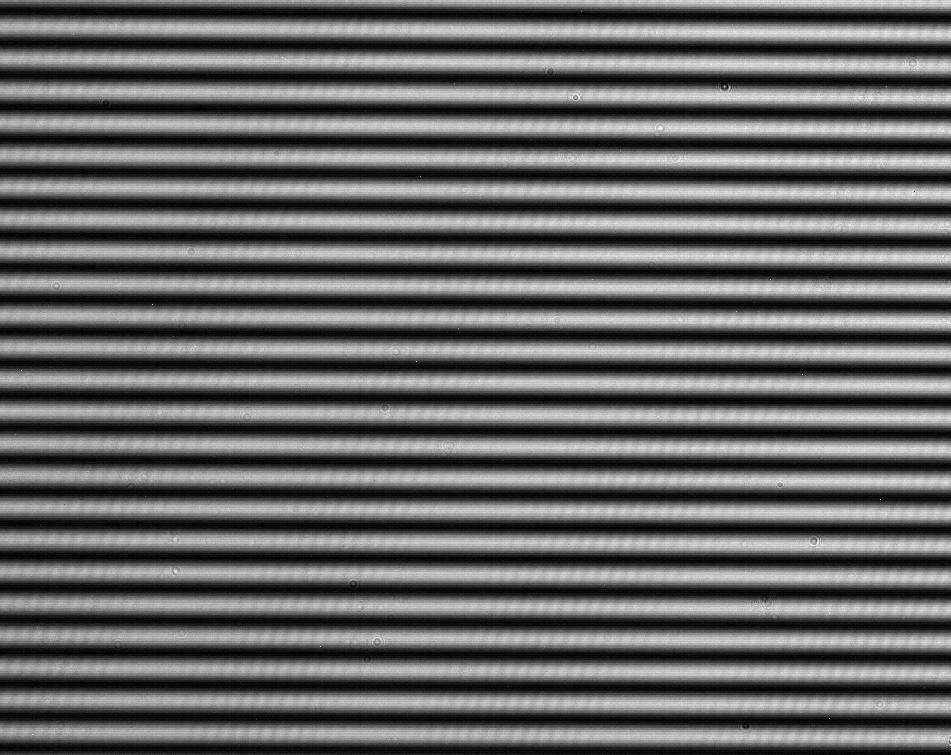}
    \includegraphics[width=0.45\textwidth]{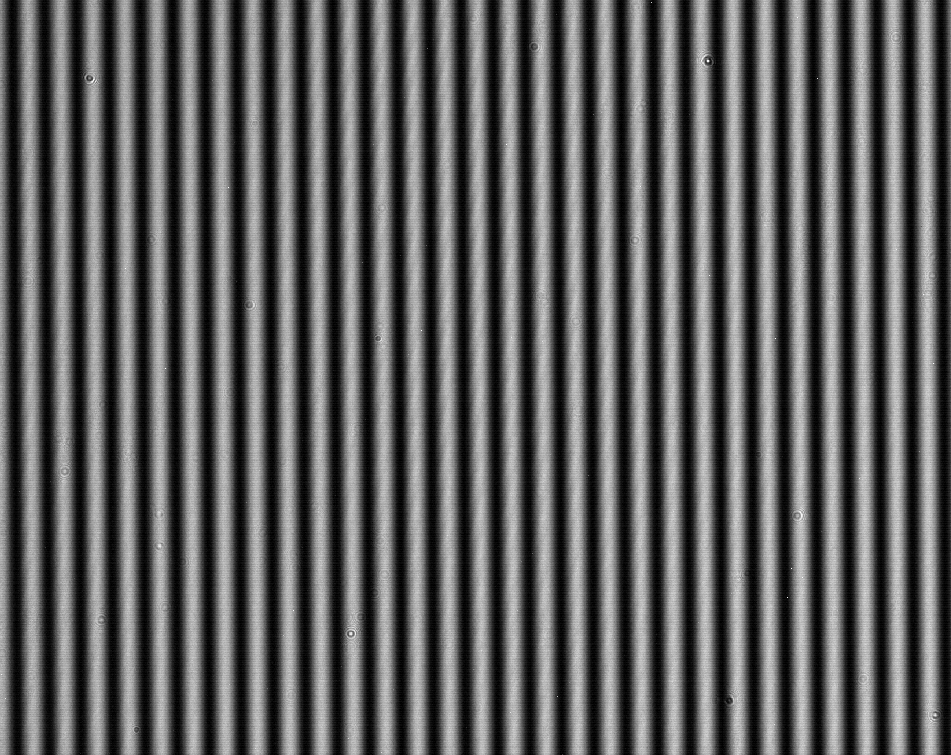}
    \includegraphics[width=0.45\textwidth]{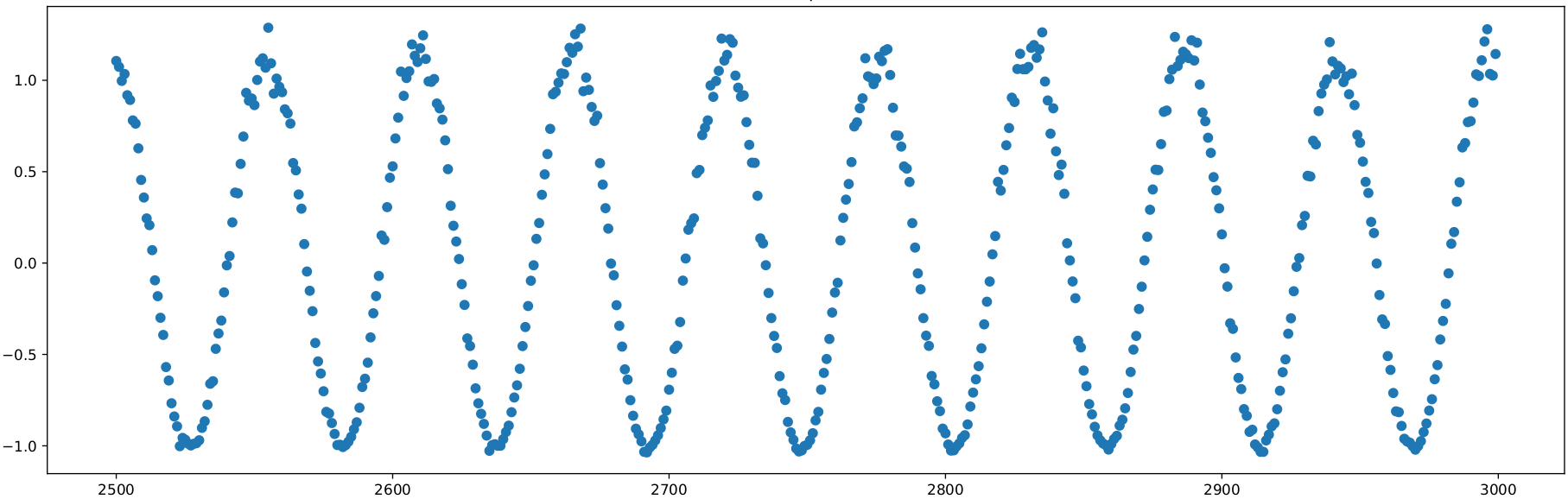}
    \includegraphics[width=0.45\textwidth]{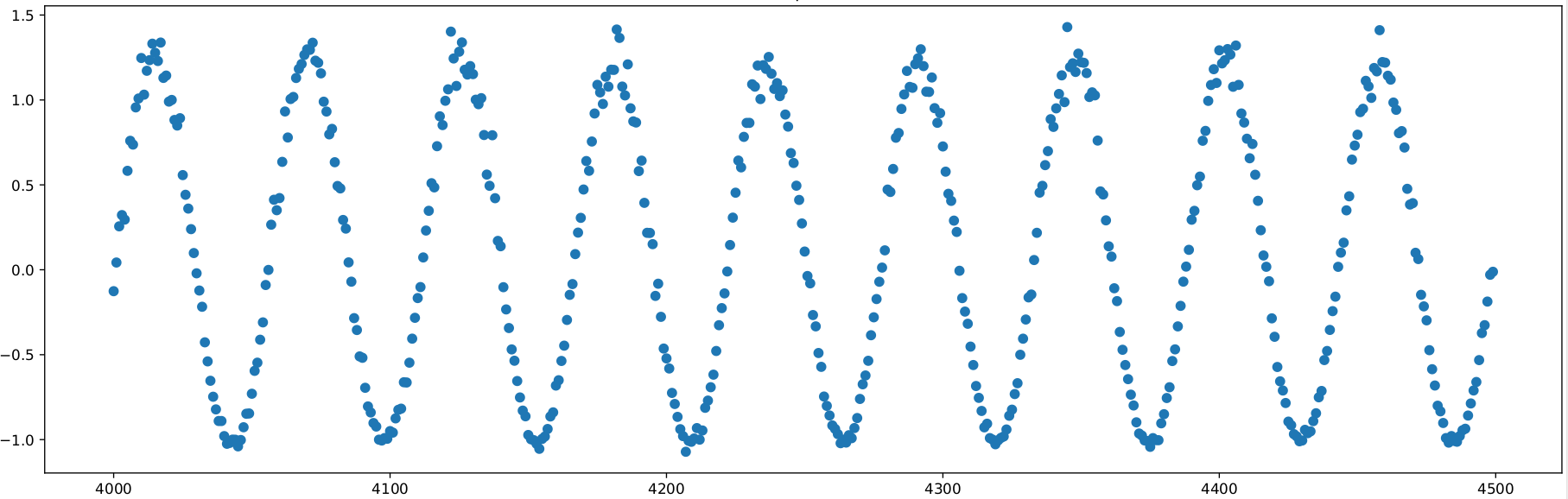}
    \caption{Interferometric fringes from the IPAG testbed pictured by the 46Mpx, top: zoom on images from the 46M gigapyx; bottom: spatial cuts. The wavelength is $632$~nm and the fibers are separated by 1 mm.}
    \label{fig:franges}
\end{figure*}

An example of fringes recorded by the gigapyx in the testbed at IPAG is showed in Fig.~\ref{fig:franges}. Both the vertical and the horizontal modes are pictured, and spatial cuts are extracted from them.\\

\section{Calibration of optical distortion}

The optical components of the telescope induce distortion. The shift in the observed positions of the stars compared to the undistorted positions may be around hundreds of pixels which is much larger than the $5.10^{-5}$~px signal we are trying to detect. We chose to calibrate the optical distortion thanks to a 2D-polynomial model called $T$.\\

Recent work on telescope stability by \cite{2022SPIE12180E..1FM} has shown that the reference stars within the telescope's FOV can serve as metrology sources in order to fit the polynomial model and then to compute the distortion function. This section presents an experimental setup that demonstrates the calibration principle of optical distortion.\\

\subsection{Numerical simulations}
\label{Numerical_simulations}

The distorted and undistorted positions of the stars can be simulated numerically to recreate a configuration similar to the one in the sky, the Zeemax software is used to simulate the distortion of a Korsh. The field distortion function is fitted by a 2D polynomial thanks to the reference stars in the FOV known from the Gaia catalog. Here the positions of reference stars are supposed to be known precisely. Finally the fitted function $T$ is applied to the distorted target star to recover its undistorted position, for more details see \cite{10.1117/12.3072791}.\\

A precision of $5.10^{-6}$ px can be reached with about $100$ stars and an $8^{th}$ order polynomial under ideal conditions (no noises). Improvements are needed to confirm these results under more realistic conditions.

\subsection{Laboratory results}

\begin{figure*}[!h]
    \centering
    \includegraphics[width=0.9\textwidth]{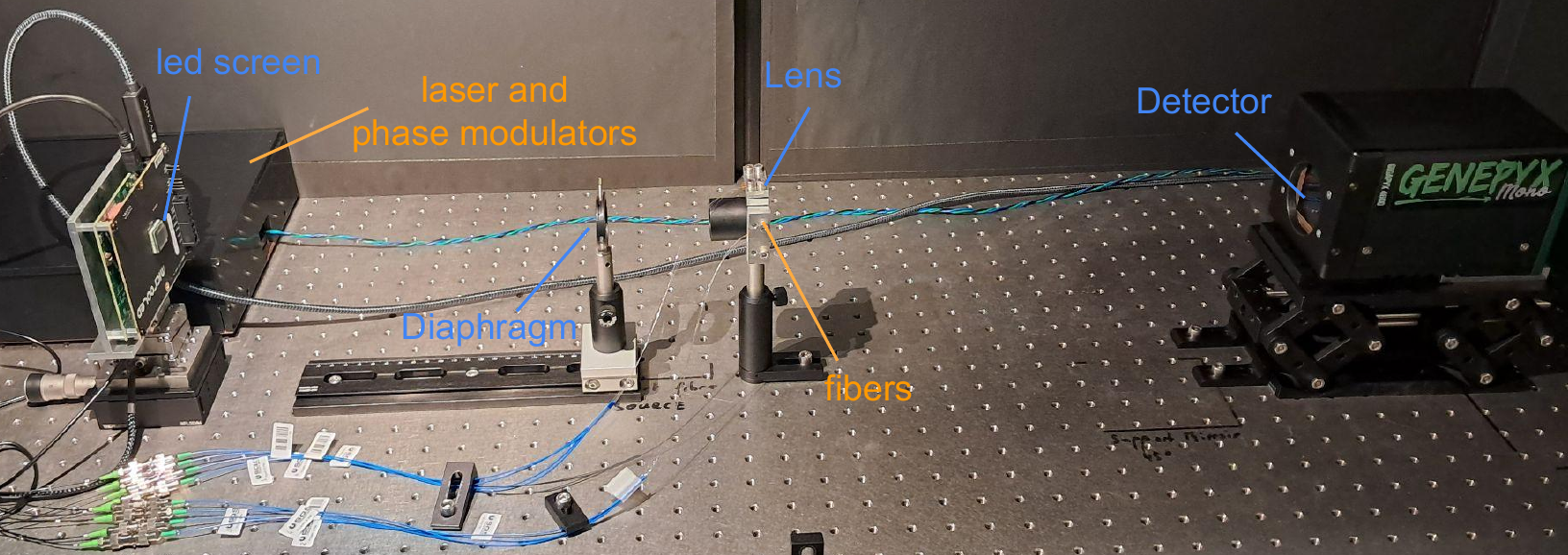}
    \caption{Photo of the distortion calibration testbed at IPAG, France (see \cite{pancher2024laboratorycharacterisationbenchhigh} for more details)}
    \label{fig:photo_banc_calib_legend}
\end{figure*}

The objective of this laboratory work is to show that the previously described method is able to calibrate a real distortion, for that purpose a testbed was developed at IPAG (France). The sky is simulated by a LED screen where each pseudo-star is one lighted pixel. A lens is placed to form an image of the screen on the detector, this is a $2f-2f$ configuration. A diaphragm is added to induce distortion. Finally, the detector captures the image, see Fig.~\ref{fig:photo_banc_calib_legend}.\\

\begin{figure*}[!h]
    \centering
    \includegraphics[width=\columnwidth]{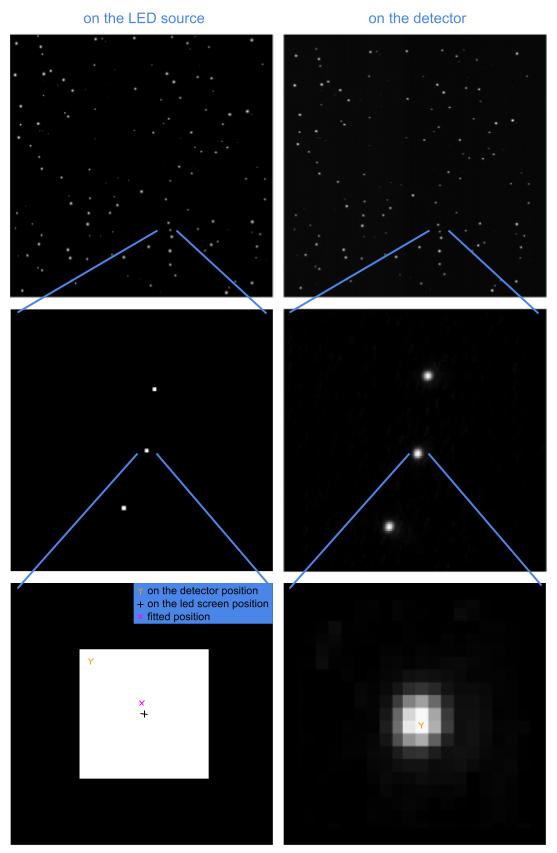}
    \caption{Optical distortion calibration, zoom from the top to the bottom. Left : pattern lighted by the LED screen, each pseudo-star is a lighted pixel; right : image taken by the gigapyx detector, each pseudo-star is the barycenter of the point spread function of the optical system. The two images seem to be similar but the right one is distorted compared to the left one.}
    \label{fig:dessin_manip_calib_disto}
\end{figure*}

The first step in the analysis is to detect the pseudo-stars in the recorded image and match them with the known positions of the lighted pixels on the LED screen. Once a match is found between the undistorted positions (LED screen) and the distorted positions (detector), the procedure described in section~\ref{Numerical_simulations} can be used to recover the distortion function. Next, the polynomial is applied to the distorted image to estimate the undistorted positions (called "fitted positions"), see Fig~\ref{fig:dessin_manip_calib_disto}. The error is calculated by computing the distance between the real undistorted positions (LED screen) and the estimated undistorted positions (by applying polynomial model). We achieved 0.04 px as a preliminary result (the distortion shifts the positions by thousands of pixels), but this must be improved particularly by integrating the results of the interferometric calibration of the detector.\\

\section{Conclusion}

We carried out the global characterization of the 46Mpx gigapyx, an adapted CMOS for an astrometric instrument on board HWO. It is compliant with the specifications of a detector for high-precision space astrometry. We also developed an interferometric testbed at IPAG in France in order to evaluate the pixel centroid positions. Finally, an ongoing study aims to demonstrate that computing the optical distortion function using the reference stars in the FOV as metrology sources works in laboratory as well as in simulations. Preliminary tests showed that the optical distortion induced in the testbed can be corrected using a polynomial function.\\

{\bf Acknowledgements}

This work has been partially supported by the LabEx FOCUS ANR-11-LABX-0013 and the CNES agency. Manon Lizzana would like to acknowledge the support of her PhD grant from CNES and Pyxalis.

This research has made use of NASA’s Astrophysics Data System Bibliographic Services and of the CDS, Strasbourg Astronomical Observatory, France.

\bibliography{author.bib}

\end{document}